\documentclass[runningheads,a4paper]{llncs}
\usepackage[top=2.5cm, bottom=2.5cm, left=2.5cm, right=2.5cm]{geometry}
\usepackage{amsmath}
\usepackage{amssymb}
\usepackage{graphicx}
\usepackage{booktabs}
\usepackage{listings}
\usepackage{xcolor}
\usepackage{hyperref}
\usepackage{caption}
\usepackage{subcaption}

\usepackage{url}
\usepackage{color}
\usepackage{cite}
\usepackage{epsfig}

\usepackage[nomargin,inline,marginclue,draft]{fixme}
\usepackage{cleveref}
\fxsetup{status=draft}
\fxsetup{theme=color, mode=multiuser} \FXRegisterAuthor{me}{ame}{\color{red}Me}
\newcommand{\orcid}[1]{\href{https://orcid.org/#1}{\includegraphics[scale=0.02]{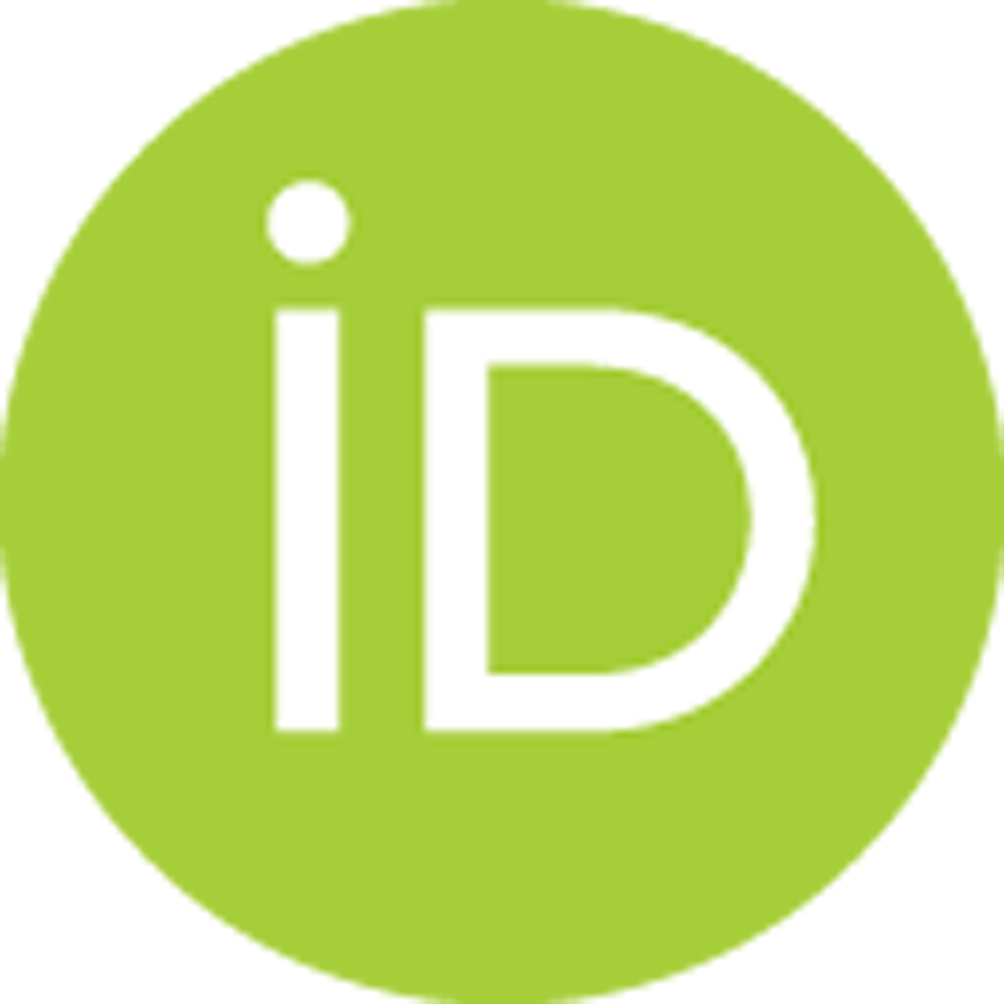}}} 

\hypersetup{
    colorlinks=true
}
\title{The Brain Tumor Segmentation (BraTS) Challenge 2023: \\ \emph{Brain MR Image Synthesis for Tumor Segmentation (BraSyn)}}
\titlerunning{BraTS 2023 Challenge}

\author{
Hongwei~Bran~Li\inst{1,2,3,\dag,\ddag} \and
Gian Marco Conte\inst{4,\dag} \and
Qingqiao Hu \inst{46,\dag} \and
Syed~Muhammad~Anwar\inst{12,13\dag} \and
Florian Kofler \inst{5,2,6,7,\dag} \and
Ivan Ezhov \inst{2, \dag} \and
Koen~van~Leemput \inst{9,\dag} \and
Marie Piraud \inst{5,\dag} \and
Maria Diaz \inst{24,\dag} \and
Byrone Cole \inst{24,\dag} \and
Evan~Calabrese \inst{27,43,\dag,\ddag} \and
Jeff~Rudie \inst{47,43,\dag,\ddag} \and
Felix~Meissen \inst{2, \dag} \and
Maruf Adewole \inst{11,\dag} \and
Anastasia~Janas \inst{17} \and
Anahita~Fathi~Kazerooni \inst{18,14,\dag,\ddag} \and
Dominic LaBella \inst{19,\dag} \and
Ahmed~W.~Moawad \inst{22} \and
Keyvan Farahani \inst{23,\dag,\ddag} \and
James Eddy \inst{24,\dag} \and
Timothy Bergquist \inst{24,\dag} \and
Verena~Chung \inst{24,\dag} \and
Russell Takeshi Shinohara \inst{14,25,\dag} \and
Farouk Dako \inst{26,\dag} \and
Walter Wiggins \inst{27,\dag,\ddag} \and
Zachary~Reitman \inst{19,\dag,\ddag} \and
Chunhao Wang \inst{19,\dag,\ddag} \and
Xinyang Liu \inst{12,13,\dag} \and
Zhifan Jiang \inst{12,13,\dag} \and
Ariana Familiar \inst{18,\dag} \and
Elaine~Johanson \inst{30,\dag} \and
Zeke Meier \inst{31,\dag} \and
Christos Davatzikos \inst{14,15,\ddag} \and
John~Freymann \inst{32,23,\ddag} \and
Justin Kirby \inst{32,23,\ddag} \and
Michel~Bilello \inst{14,15,\ddag} \and
Hassan~M.~Fathallah-Shaykh \inst{33,\ddag} \and
Roland~Wiest \inst{34,35,\ddag} \and
Jan Kirschke \inst{21,\ddag} \and
Rivka~R.~Colen \inst{36,37,\ddag}
Aikaterini Kotrotsou \inst{37,\ddag}
Pamela~Lamontagne \inst{38,\ddag}
Daniel Marcus \inst{39,40,\ddag}
Mikhail~Milchenko \inst{39,40,\ddag}
Arash Nazeri \inst{40,\ddag}
Marc-André~Weber \inst{41,\ddag}
Abhishek~Mahajan \inst{42,\ddag}
Suyash~Mohan \inst{14,15,\ddag}
John~Mongan \inst{43,\ddag}
Christopher~Hess \inst{43,\ddag}
Soonmee~Cha \inst{43,\ddag}
Javier~Villanueva-Meyer \inst{43,\ddag} \and
Errol~Colak \inst{44,\ddag} \and
Priscila~Crivellaro \inst{44,\ddag} \and
Andras~Jakab \inst{45,\ddag} \and
Jake~Albrecht \inst{24,\dag} \and
Udunna~Anazodo \inst{29,\dag} \and
Mariam Aboian \inst{17,\dag,\ddag} \and
Thomas Yu \inst{10} \and
Verena~Chung \inst{24} \and
Timothy~Bergquist \inst{24} \and
James Eddy \inst{24} \and
Jake~Albrecht \inst{24} \and
Ujjwal~Baid \inst{14, 15, 16} \and
Spyridon Bakas \inst{14, 15, 16} \and
Marius~George~Linguraru \inst{8,\dag} \and
Bjoern~Menze \inst{1} \and
Juan~Eugenio~Iglesias \inst{10,\dag, \P} \and
Benedikt~Wiestler\inst{3, \dag, \P} 
}

\authorrunning{Li et al.}

\institute{\scriptsize{University of Zurich, Switzerland 
\and Department of Informatics, Technical University Munich, Germany 
\and Klinikum rechts der Isar, Technical University of Munich, Germany  
\and Mayo Clinic, Rochester, USA 
\and Helmholtz AI, Helmholtz Munich, Germany 
\and TranslaTUM - Central Institute for Translational Cancer Research, Technical University of Munich, Germany \and  
Department of Diagnostic and Interventional Neuroradiology, School of Medicine, Klinikum rechts der Isar, Technical University of Munich, Germany \and  
Children's National Hospital, Washington DC, USA and George Washington University, USA \and 
Finnish Center for Artificial Intelligence, Finland \and 
Athinoula A. Martinos Center for Biomedical Imaging, Harvard Medical School, USA \and 
Medical Artificial Intelligence (MAI) Lab, Crestview Radiology, Lagos, Nigeria \and  
Children's National Hospital, Washington DC, USA \and  
George Washington University, Washington DC, USA \and  
Center for AI and Data Science for Integrated Diagnostics (AI2D) I\& Center for Biomedical Image Computing and Analytics (CBICA), University of Pennsylvania, Philadelphia, PA, USA \and  
Department of Radiology, Perelman School of Medicine, University of Pennsylvania, Philadelphia, PA, USA \and  
Department of Pathology and Laboratory Medicine, Perelman School of Medicine, University of Pennsylvania, Philadelphia, PA, USA \and  
Yale University, New Haven, CT, USA \and  
Children’s Hospital of Philadelphia, University of Pennsylvania, Philadelphia, PA, USA \and  
Duke University Medical Center, Department of Radiation Oncology, USA \and  
Athinoula A Martinos Center for Biomedical Imaging, Massachusetts General Hospital, Boston, MA, USA \and  
Department of Neuroradiology, Technical University of Munich, Munich, Germany \and  
Mercy Catholic Medical Center, Darby, PA, USA \and  
Cancer Imaging Program, National Cancer Institute, National Institutes of Health, Bethesda, MD 20814, USA \and  
Sage Bionetworks, USA \and  
Center for Clinical Epidemiology and Biostatistics, University of Pennsylvania, Philadelphia, USA \and  
Center for Global Health, Perelman School of Medicine, University of Pennsylvania, Philadelphia, Pennsylvania, USA \and  
Duke University Medical Center, Department of Radiology, USA \and  
Department of Applied Mathematics and Computer Science, Technical University of Denmark, Denmark \and  
Montreal Neurological Institute (MNI), McGill University, Montreal, QC, Canada \and  
PrecisionFDA, U.S. Food and Drug Administration, Silver Spring, MD, USA \and  
Booz Allen Hamilton, McLean, VA, USA \and  
Leidos Biomedical Research, Inc., Frederick National Laboratory for Cancer Research, Frederick, MD 21701, USA \and  
Department of Neurology, The University of Alabama at Birmingham, Birmingham, AL, USA \and  
Institute for Surgical Technology and Biomechanics, University of Bern, Bern, Switzerland \and  
Support Centre for Advanced Neuroimaging Inselspital, Institute for Diagnostic and Interventional Neuroradiology, Bern University Hospital, Bern, Switzerland \and  
University of Pittsburgh Medical Center, PA, USA \and  
Department of Diagnostic Radiology, University of Texas MD Anderson Cancer Center, Houston, TX, USA \and  
Department of Psychology, Washington University, St. Louis, MO, USA \and  
Neuroimaging Informatics and Analysis Center, Washington University, St. Louis, MO, USA \and  
Department of Radiology, Washington University, St. Louis, MO, USA \and  
Institute of Diagnostic and Interventional Radiology, Pediatric Radiology and Neuroradiology, University Medical Center Rostock, Ernst-Heydemann-Str. 6, 18057 Rostock, Germany \and  
Tata Memorial Centre, Homi Bhabha National Institute, Mumbai, India \and  
University of California San Francisco, CA, USA \and  
University of Toronto, Toronto, ON, Canada \and  
University of Debrecen, Hungary  
Stony Brook University, NY, USA  
}
\\
\textsuperscript{\dag} People involved in the organization of the challenge.\\
\textsuperscript{\ddag} People contributing data from their institutions.\\
\textsuperscript{\S} People involved in annotation process.\\ 
\textsuperscript{\P} Equal senior authors.\\ 
\textsuperscript{**} Corresponding author: \email{\{hongwei.li@uzh.ch\}}}

\begin{document}
    \mainmatter
    \maketitle
    \setcounter{footnote}{0} 
    \begin{abstract}
Automated brain tumor segmentation methods have become well-established and reached performance levels offering clear clinical utility. These methods typically rely on four input magnetic resonance imaging (MRI) modalities: T1-weighted images with and without contrast enhancement, T2-weighted images, and FLAIR images. However, some sequences are often missing in clinical practice due to time constraints or image artifacts, such as patient motion. Consequently, the ability to substitute missing modalities and gain segmentation performance is highly desirable and necessary for the broader adoption of these algorithms in the clinical routine.
In this work, we present the establishment of the Brain MR Image Synthesis Benchmark (BraSyn) in conjunction with the Medical Image Computing and Computer-Assisted Intervention (MICCAI) 2023. The primary objective of this challenge is to evaluate image synthesis methods that can realistically generate missing MRI modalities when multiple available images are provided. The ultimate aim is to facilitate automated brain tumor segmentation pipelines. The image dataset used in the benchmark is diverse and multi-modal, created through collaboration with various hospitals and research institutions. 

    \end{abstract}
    
    
    \section{Introduction}
Manual segmentation of brain tumors in magnetic resonance images (MRI) is a tedious task with high variability among raters. However, for the objective assessment of tumor response (as outlined in the RANO criteria \cite{wen2010updated}, reliable volumetry of tumors are essential. A recent study showed that AI-based decision support through automated tumor segmentation clearly benefits even expert clinicians in this task \cite{vollmuth2023artificial}. Many recent works have developed automated segmentation methods to address this using deep learning (DL) \cite{kamnitsas2017efficient,pereira2016brain,wang2019automatic}. These algorithms mostly require four input MRI modalities, typically T1-weighted (T1-w) images with and without contrast enhancement, T2-weighted (T2-w) images, and FLAIR images during the inference stage. However, a common challenge in clinical routines is missing MR sequences, e.g., because of time constraints and/or image artifacts caused by patient motion. Some sequences, especially FLAIR and T1, are often missing from routine MRI examinations \cite{conte2021generative}. Therefore, the synthesis of missing modalities is desirable and necessary for the more widespread use of such algorithms in clinical routine.

This challenge calls for modality synthesis algorithms of MRI volumes, enabling a straightforward application of \emph{BraTS} routines in centers with less extensive imaging protocols or for analyzing legacy datasets. As BraTS focuses on brain tumor image analysis, this modality synthesis task will enable the application of the downstream image segmentation routines even in incomplete datasets. 

Generating missing MRI sequences holds promise to facilitate image segmentation and has attracted growing attention in recent years \cite{conte2021generative,iglesias2013synthesizing,anwar2018medical}. For example, deep learning networks based on generative adversarial networks (GANs) have been explored for this task with promising results \cite{li2019diamondgan,thomas2022improving,iglesias2021joint}. From a technical standpoint, these algorithms need to overcome a multitude of challenges: First, the image resolutions of the individual sequences might differ; for example, FLAIR images tend to be acquired using 2D sequences, leading to anisotropic resolution, and thus matching the resolution of other 3D imaging sequences only poorly. Second, motion artifacts may be present in some of the sequences. At the same time, MRI bias fields may differ in their local impact on the different image modalities, leading to spatially inconstant artifacts. And third, a general domain shift between the training and test sets due to different acquisition settings and types of scanners can be expected in almost any large and multi-institutional dataset \cite{hu2022domain}. All these effects must be considered when developing methods for synthesizing MR images. Questions about how to deal with these challenges, for example, by choosing adequate metrics or invariance properties of the algorithms and network architecture, have yet to be answered.

In previous BraTS challenges, we have set up publicly available datasets – and algorithms – for multi-modal brain glioma segmentation \cite{baid2021rsna,bakas2018identifying,menze2014multimodal}. In our whole-volume MRI synthesis task, we will build on these efforts (and the previously generated data sets) to further the development of much-needed computational tools for data integration and homogenization. This new challenge will enable a broader application of the tumor segmentation algorithms developed in previous BraTS editions (that require a fixed set of image modalities) and better integration with other downstream routines used for quantitative neuro-image analysis (that only work well for brain images without perturbations from artifacts or lesion). The resulting MRI synthesis is essential to develop effective, generalizable, reproducible methods for analyzing high-resolution MRI of brain tumors. \emph{BraSyn} will include well-established data from multiple sites from previous BraTS challenges, adding new inference tasks beyond image segmentation. Resulting algorithms will have the potential to benefit automated brain (tumor) image processing and improve the clinical risk stratification tools for early interventions, treatments, and care management decisions across hospitals and research institutions worldwide.

    \section{Materials}
        \subsection{Dataset}

The \emph{BraSyn-2023} dataset is based on the \emph{RSNA-ASNR-MICCAI BraTS 2021} dataset \cite{baid2021rsna} and involves the retrospective collection of multi-parametric MRI (mpMRI) scans of brain tumors from various institutions. These scans were acquired under standard clinical conditions but with different imaging equipment and protocols, resulting in a wide range of image quality due to variations in clinical practices across institutions. Expert neuroradiologists meticulously reviewed and approved ground truth annotations for each tumor subregion. The annotated tumor subregions are based on observed features visible to trained radiologists (referred to as VASARI features) and include the Gd-enhancing tumor (ET - labeled as 4), peritumoral edematous/infiltrated tissue (ED - labeled as 2), and necrotic tumor core (NCR - labeled as 1). The ET represents the enhancing part of the tumor, characterized by areas with both strong and weak enhancement on T1Gd MRI. The NCR represents the necrotic core of the tumor, which appears hypointense on T1Gd MRI. The ED refers to the peritumoral edematous and infiltrated tissue, identified by the abnormal hyperintense signal observed on the T2 FLAIR volumes, encompassing both the non-enhancing infiltrative tumor and vasogenic edema in the peritumoral region.

    \begin{figure}[t]
          \centering
    \includegraphics[width=0.8\linewidth]{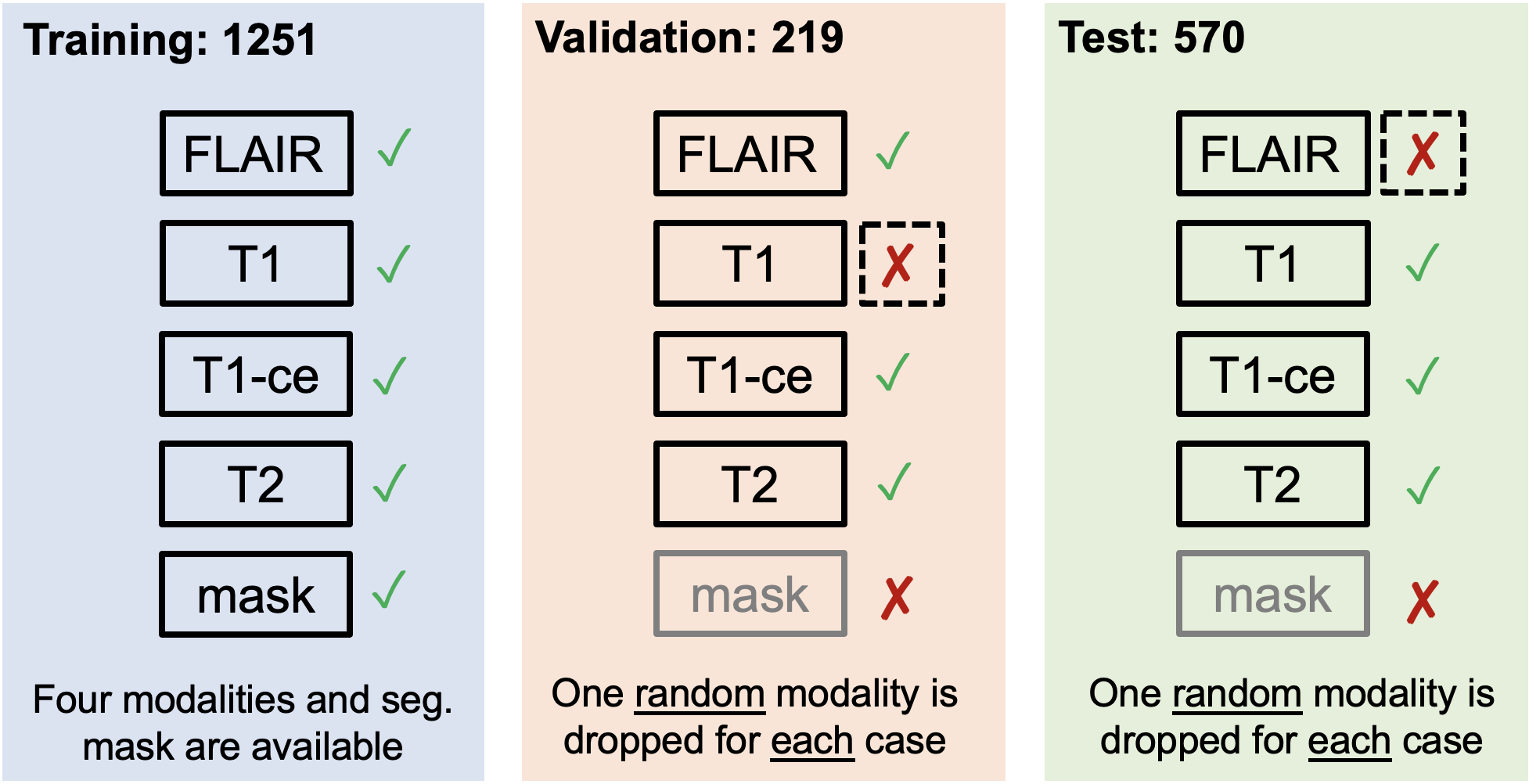}  
          \caption{{The details of training, validation, and test sets.} During the validation and test stages, for each subject, the segmentation mask corresponding to images is not available, and one of the four modalities will be \textbf{randomly} excluded (`dropout'). The participants are required to synthesize any missing modalities in the test stage.}
        \label{main}
    \end{figure}
 
In line with the approach of algorithmic evaluation in machine learning, the data utilized in the \emph{BraSyn-2023} challenge is partitioned into training, validation, and private testing datasets. For the training data, participants are provided with four complete image modalities along with their respective segmentation labels, following a setup akin to the segmentation challenge.
In the validation and test sets, a single modality out of the four sequences will be randomly omitted for each case. This deliberate omission aims to assess the effectiveness and performance of the submitted image synthesis methods.

\label{sec:data}

\noindent
{\textbf{Imaging data description.}} 
The mpMRI scans included in the BraTS 2023 challenge contain a) pre-contrast (T1) and b) post-contrast T1-weighted (T1Gd (Gadolinium)), c) T2-weighted (T2), and d) T2 Fluid Attenuated Inversion Recovery (T2-FLAIR) volumes, acquired with different protocols and various scanners from multiple institutions. 
    
All the BraTS mpMRI scans have undergone standardized pre-processing, which includes the conversion of DICOM files to the NIfTI file format\cite{nifti}, co-registration to the same anatomical template (SRI24) \cite{SRI_rohlfing2010sri24}, resampling to a uniform isotropic resolution ($1mm^{3}$), and finally skull-stripping. The pre-processing pipeline is publicly available through the Cancer Imaging Phenomics Toolkit (CaPTk) \footnote{\url{https://cbica.github.io/CaPTk/}} \cite{captk,captk_2,captk_3} and Federated Tumor Segmentation (FeTS) tool \footnote{\url{https://github.com/FETS-AI/Front-End/}}. Conversion to NIfTI strips the accompanying metadata from the DICOM images and essentially removes all Protected Health Information (PHI) from the DICOM headers. Furthermore, skull-stripping mitigates potential facial reconstruction/recognition of the patient \cite{NEJMc1908881,NEJMc1915674}. The specific approach we have used for skull stripping is based on a novel DL approach that accounts for the brain shape prior and is agnostic to the MRI sequence input \cite{thakur2020brain}.
         
All imaging volumes have then been segmented using the STAPLE \cite{warfield2004simultaneous} fusion of previous top-ranked BraTS algorithms, namely, nnU-Net \cite{isensee2020nnu}, DeepScan \cite{mckinley2018ensembles}, and DeepMedic \cite{kamnitsas2017efficient}. Subsequently, volunteer neuro-radiologists with different levels of rank and experience undertook the manual refinement of these fused labels, adhering to a consistently communicated annotation protocol. The refined annotations, after this manual refinement process, were ultimately reviewed and approved by highly experienced board-certified attending neuro-radiologists who possess over 15 years of expertise in glioma-related work

\subsection{Evaluation Metrics}  

The inference task that submitted algorithms have to solve is the following: When presented with a test set, one of the four modalities will be missing. The algorithm must predict a plausible brain tumor image for the missing modality. The image will be evaluated in terms of general image quality as well as by the performance of a downstream tumor segmentation algorithm that will be applied to the completed image set. For ranking the contributions, we will be using the following set of metrics: First, we calculate the structural similarity index measure (SSIM) as a direct image quality metric to quantify how realistic the synthesized images are compared to clinically acquired real images. SSIMs will be calculated in the tumor area and in the healthy part of the brain, resulting in two scores for each test subject. Second, we will evaluate whether the synthesized images are helpful for a segmentation algorithm. To this end, we will segment the four modalities including a synthesized volume with a state-of-the-art BraTS segmentation algorithm, and calculate Dice scores for three tumor structures as `indirect' metrics. The automated segmentation will be performed by the final \emph{FeTS} algorithm \cite{pati2022federated}, leveraging the model pre-trained in the FETS brain tumor segmentation initiative. Participants will have access to the algorithm to fine-tune their methods and optimize their performance for our evaluation scenario. To assess image quality and segmentation accuracy, a total of five ranking metrics will be utilized.

 
\subsection{Ranking strategy}  
For the final ranking of the participants, an equally weighted rank-sum is computed for each case of the test set, considering all the aforementioned metrics. This will rank algorithms according to each metric and sum up all ranks. The synthesis task will have five ranking scores: three Dice scores for each tumor tissue and two SSIMs for the image quality of tumor and non-tumor regions. The participating team with the best rank-sum will win each challenge task. 


\subsection{Participation Policy and Timeline}

The challenge will begin by releasing a training dataset containing imaging data and corresponding labels for ground truth. Participants are allowed to start developing and training their methods using this dataset.

After three weeks, the validation data will be released, allowing participants to obtain initial results on unseen data. They can include these results in their submitted short MICCAI LNCS papers, along with their cross-validated results on the training data. While the ground truth for the validation data will not be disclosed, participants can make multiple submissions on the online evaluation platforms. The top-performing teams in the validation phase will be invited to prepare slides for a brief oral presentation of their methods during the BraTS challenge at MICCAI 2023.

During training, participants can utilize publicly available brain scans from healthy individuals. However, to ensure a fair comparison among methods, participants must explicitly mention any additional data in their submitted manuscripts and report results using only the training data provided by the organizers, focusing on potential differences in outcomes.

Finally, all participants will be assessed and ranked based on an unseen testing dataset, which will not be accessible once they have uploaded their containerized method to the evaluation platforms. The final rankings and winners will be announced at MICCAI 2023, with the top-ranked teams receiving monetary prizes.

    \section*{Acknowledgments}
 We thank all the data contributors, annotators, and approvers for their time and efforts. This work is supported by NIH funds 5-U01-CA-242871-03 (B.H. and S.B.) and 5-R01-CA-269948-02 (B.W. and S.B.). 

    
    \section*{Funding}
    
    Research reported in this publication was partly supported by the National Institutes of Health (NIH) under award numbers: NIH/NCI/ITCR:U01CA242871. The content of this publication is solely the responsibility of the authors and does not represent the official views of the NIH.

    \bibliographystyle{ieeetr}
    \bibliography{bibliography.bib}
    \newpage
    \appendix
\end{document}